\documentclass[a4paper,12pt,oneside]{article}
\usepackage{graphicx}
\usepackage[all]{xy}
\usepackage[centertags]{amsmath}
\usepackage{latexsym}

\usepackage{amsfonts}
\usepackage{amssymb,amsthm}
\usepackage{indentfirst}
\usepackage[english]{babel}
\usepackage[latin1]{inputenc}
\usepackage{eufrak}

\usepackage{amsfonts}

\usepackage{amssymb,amsthm}
\usepackage[english]{babel}

\usepackage{newlfont}

\newtheorem{remark}{Remark}[section]

\theoremstyle{plain}
\newtheorem{lemma}{Lemma}[section]
\newtheorem{proposition}{Proposition}[section]
\newtheorem{theorem}{Theorem}[section]
\newtheorem{definition}{Definition}[section]
\newtheorem{corollary}{Corollary}[section]

\newtheorem{example}{Example}[section]

\newcommand{\R}{\mathbb R}
\newcommand{\C}{\mathbb C}

\newcommand{\beqn}{\begin{eqnarray}}
\newcommand{\eeqn}{\end{eqnarray}}
\newcommand{\beq}{\begin{eqnarray}}
\newcommand{\eeq}{\end{eqnarray}}
\newcommand{\bpro}{\begin{proposition}}
\newcommand{\epro}{\end{proposition}}
\newcommand{\blem}{\begin{lemma}}
\newcommand{\elem}{\end{lemma}}
\newcommand{\bdfn}{\begin{definition}}
\newcommand{\edfn}{\end{definition}}
\newcommand{\bcor}{\begin{corollary}}
\newcommand{\ecor}{\end{corollary}}
\newcommand{\bthm}{\begin{theorem}}
\newcommand{\ethm}{\end{theorem}}
\newcommand{\bex}{\begin{example}}
\newcommand{\eex}{\end{example}}
\newcommand{\brmq}{\begin{remark}}
\newcommand{\ermq}{\end{remark}}
\newcommand{\benum}{\begin{enumerate}}
\newcommand{\eenum}{\end{enumerate}}
\newcommand{\bitem}{\begin{itemize}}
\newcommand{\eitem}{\end{itemize}}

\theoremstyle{plain}

\usepackage[latin1]{inputenc}

\title{Cloaking via change of variables in elastic impedance tomography}

 \author{ Andr\'e Diatta and S\'ebastien Guenneau \footnote{ \footnotesize Aix-Marseille Universit\'e, CNRS, Institut Fresnel, Ecole Centrale Marseille, 13013 Marseille, France
\newline
E-mail: andre.diatta@fresnel.fr; sebastien.guenneau@fresnel.fr}
 }

\begin{document}

\maketitle

\begin{abstract}We discuss the concept of cloaking for elastic impedance tomography, in which, we seek information on the elasticity tensor of an elastic medium from the knowledge of  measurements on its boundary. We derive some theoretical  results illustrated by some numerical simulations.

\end{abstract}

\section{Introduction}
A number of papers have studied the properties of the transformed Navier equations \cite{willis,norris,brun} in the context of cloaking
\cite{pendry,leonhardt} in the past few years. Such mathematical models are known from researchers working in the context of
electric impedance tomography \cite{kohn1,lee,greenleaf,kohn2}, wherein a region of space is cloaked with respect to
sensing if its contents, and the surrounding cloak, are inaccessible to conductivity measurements. Fortunately, the
conductivity equation is isomorphic to the acoustic equation, hence acoustic cloaking \cite{cummer,chan} in the
context of tomography can be straightforwardly envisioned. However, for medical imaging applications \cite{fink}
whereby shear and pressure waves are inherently coupled when propagating within a human body,
we need to translate the work of \cite{kohn1,lee,greenleaf,kohn2} in the language of elastic impedance tomography.
In a homogeneous  linear elastic material with fourth order constitutive tensor  $\C,$  the
partial differential equation (PDE) for elastostatics, reads
\begin{eqnarray}\label{eq:navier-general}
{\mathbf
 \nabla}. \left(\C: (\nabla {\mathbf u})\right)= \nabla.\sigma={\mathbf 0,}
\end{eqnarray}
where $\sigma:=\C: \nabla {\mathbf u}$ is the stress tensor. This relates the displacement field ${\mathbf u}$ and its gradient to the  stress tensor $\sigma.$ From boundary measurement, the  PDE  (\ref{eq:navier-general}) gives rise to the  Dirichlet-to-Neumann map
\begin{eqnarray}\label{eq:DtNmap}
{\mathbf
\Lambda_\C} :{\mathbf  u}|_{\partial\Omega} \mapsto  \left(\C: \nabla {\mathbf u}\right).{\mathbf n}|_{\partial\Omega}
\end{eqnarray}
which  relates the displacement field  to the gradient traction on the boundary ${\partial\Omega}$ of a bounded domain ${\partial\Omega}$ in $\R^n.$  Here and throughout this note,  ${\mathbf n}$  is the outward unit normal to $\partial \Omega.$ The results within  this paper apply to any bounded (open) domain in $\mathbb R^n$, $n\ge 2,$ but for simplicity and without lost of generality, we will take ${\Omega}$  to be a ball  when $n\ge 3$ (resp. a disc, for $n=2)$  of radius $R$, so its boundary is  the  sphere (resp. circle) ${\partial\Omega}=S({\mathbf 0},R)$.

\section{Main theoretical results}

Elastic impedance tomography seeks information on the elasticity tensor $\C$ of a medium occupying a region $\Omega,$ from the knowledge of  ${\mathbf  u}$ on the boundary ${\partial\Omega}$ of $\Omega.$

In an isotropic homogeneous elastic media, the general Hooke's law reads
\begin{eqnarray}\label{T-isotropic}
{ \mathbf \sigma}={ \mathbf \sigma_1}=\lambda {\mathbf  I} (\nabla.{\mathbf u})+\mu (\nabla{\mathbf u}+(\nabla{\mathbf u})^T)
\end{eqnarray}
where $\lambda$ and $\mu$ are Lam\'e constants (e.g.we could have  $\lambda=2.3$ and $\mu=1$  for a Poisson ratio close to $0.35$), and ${\mathbf I}$ is the second order unit tensor.

\subsection{The construction of  $  (\C:\nabla  {\mathbf u}).  {\mathbf n}  $}
Let $\Omega$ be a bounded open subset of  $\mathbb R^n,$ $n=2$ or $3.$  We let  $ H^{{1}}(\Omega)$ stand for the Hilbert space of vector fields ${\mathbf w}$ in $\Omega$ such that ${\mathbf w}\in (L^2(\Omega))^n$  and $\nabla{\mathbf w}\in (L^2(\Omega))^{n\times n}.$  We denote by  $ H^{\frac{1}{2}}(\partial \Omega),$ the space of vector fields ${\mathbf f}$ on the boundary $\partial \Omega$ of $ \Omega,$ which are traces ${\mathbf w}|_{\partial \Omega}={\mathbf f}$ of some 
  element  ${\mathbf w}$ of $ H^{{1}}(\Omega).$ 
 
For  a vector field ${\mathbf u}$ such that
$ \C:\nabla  {\mathbf u}\in (L^2(\Omega))^{n\times n}$ and  $\nabla.(\C:\nabla  {\mathbf u})\in (L^2(\Omega) )^n,$ we define the linear functional 
\begin{eqnarray}\label{def:Xu}
X_ {\mathbf u}({\mathbf f}):=\displaystyle\int_{\Omega} [ (\C:\nabla  {\mathbf u}):\nabla {\mathbf w}  +(\nabla.(\C:\nabla  {\mathbf u})) .\mathbf w] d{\mathbf x},
\end{eqnarray}
for every ${\mathbf f}\in H^{\frac{1}{2}}(\partial \Omega),$
 where ${\mathbf w}$ is some element of $ H^{{1}}(\Omega)$ satisfying ${\mathbf w}|_{\partial \Omega}={\mathbf f}.$
As can be seen easily,  $X_ {\mathbf u}$ is well defined, for the right-hand side of (\ref{def:Xu}) is the same irrespective whether we use ${\mathbf w}$ or ${\mathbf w'}$  to write it, as long as ${\mathbf w}$ and ${\mathbf w'}$ coincide on the boundary $\partial \Omega.$
 If $\C$ is an element of $(L^{\infty}(\Omega))^{n^4}$ and ${\mathbf u}$ is  smooth enough,  then $X_ {\mathbf u}$ is simply  
\begin{eqnarray}\label{eq2:Xu}
X_ {\mathbf u}({\mathbf f})
= \displaystyle \int_{\partial \Omega}   \left(  (\C:\nabla  {\mathbf u}).  {\mathbf n} \right).{\mathbf{ w}}  ~d{{\mathbf
x}}
\end{eqnarray}
Now from Cauchy-Schwarz' inequality,  there is some constant $C_0>0$ such that $|X_ {\mathbf u}({\mathbf f})|\leq C_0||{\mathbf f}||_{ H^{\frac{1}{2}}(\partial\Omega)}.$ Hence, $X_ {\mathbf u}$ is continuous.
 Thus   $X_ {\mathbf u}$ is an element of the dual $  H^{-\frac{1}{2}}(\partial\Omega)$  of the space $H^{\frac{1}{2}}(\partial\Omega).$ From the Riesz representation theorem, there exists some
 ${\mathbf f_0}\in H^{\frac{1}{2}}(\partial\Omega)$ such that $ X_ {\mathbf u}({\mathbf f})=\langle{\mathbf f_0},{\mathbf f}\rangle_{H^{\frac{1}{2}}(\partial \Omega)}$
 for every  ${\mathbf f}\in  H^{\frac{1}{2}}(\partial\Omega).$ 
 
By definition, we write $ \left(\C: \nabla {\mathbf u}\right).{\mathbf n}|_{\partial\Omega}:= {\mathbf f_0}. $ We  have proved the following 
 
\begin{lemma}\label{XX} For  any vector field ${\mathbf u}$ such that
$ \C:\nabla  {\mathbf u}\in (L^2(\Omega))^{n\times n}$ and  $\nabla.(\C:\nabla  {\mathbf u})\in (L^2(\Omega) )^n,$ the trace $ \left(\C: \nabla {\mathbf u}\right).{\mathbf n}|_{\partial\Omega}$ exists and is in $H^{\frac{1}{2}}(\partial\Omega).$ \end{lemma}
Notice that solutions ${\mathbf u}$  of Eq. (\ref{eq:navier-general}) only need to satisfy $ \C:\nabla  {\mathbf u}\in (L^2(\Omega))^{n\times n}.$  The definition of $ \left(\C: \nabla {\mathbf u}\right).{\mathbf n}|_{\partial\Omega}$ does not depend on  the choice of ${\mathbf u},$ but only on the trace ${\mathbf f}={\mathbf u}|_{\partial \Omega},$ we simply write   $\Lambda_\C({\mathbf f})=\left(\C: \nabla {\mathbf u}\right).{\mathbf n}|_{\partial\Omega}.$

\subsection{The invariance by change of variable}\label{chap:invariance-change-of-coordinates}
If ${\mathbf M}$ and ${\mathbf  N}$ are two $n\times n$ matrix  (second order tensor) fields, with $n=2,$ or $3,$ we set  
$\langle {\mathbf M}, {\mathbf  N}\rangle:=\text{Trace} \left(  {\mathbf M} {\mathbf N}^T\right),$

so that 
\begin{eqnarray}
\langle \C:\nabla {\mathbf u},\nabla  {\mathbf v}\rangle=\text{Trace}\left( (\C:\nabla {\mathbf u})(\nabla  {\mathbf v})^T\right)=\displaystyle\sum_{i} \left( (\C:\nabla {\mathbf u})(\nabla  {\mathbf v})^T\right)_{ii}= (\C:\nabla {\mathbf u}):\nabla  {\mathbf v}.
\end{eqnarray} We denote by $ \EuFrak{a_\C}$ the following bilinear form acting on vector fields
\begin{eqnarray}
 \EuFrak{a_\C} ({\mathbf u},{\mathbf v})=\displaystyle \int_\Omega \langle \C:\nabla {\mathbf u},\nabla{\mathbf v}\rangle d{\mathbf
x}=\displaystyle \int_\Omega ( \C:\nabla {\mathbf u}):\nabla{\mathbf v} ~d{\mathbf x}
\end{eqnarray}
Using integration by parts, we have
\begin{eqnarray}\label{eq:DtNmap4}
 \EuFrak{a_\C} ({\mathbf u},{\mathbf u})
=  \displaystyle \int_{\partial \Omega}   \left(  (\C:\nabla  {\mathbf u}).  {\mathbf n} \right).{\mathbf{ u}}  ~d{{\mathbf
x}}-  
 \displaystyle \int_{ \Omega}  \left( \nabla . {\mathbf
(\C:\nabla {\mathbf u}) } \right).{\mathbf u} ~ d{\mathbf
x}
\end{eqnarray}

So, if  $\tilde{\mathbf u}$ is a solution of  Eq. (\ref{eq:navier-general}), we then have
$  \EuFrak{a_\C} ( \tilde{\mathbf u}, \tilde{\mathbf u}) 
=
 \displaystyle \int_{\partial \Omega}   \left(  (\C:\nabla \tilde {\mathbf u}).  {\mathbf n} \right).\tilde{\mathbf{ u}}  ~d{{\mathbf
x}}.
$
If, in addition,  ${ \tilde{ \mathbf  u}}|_{\partial\Omega}={\mathbf f},$  we end up having 
\begin{eqnarray}\label{eq:DtNmap111}
 \EuFrak{a_\C} ( \tilde{\mathbf u}, \tilde{\mathbf u}) =
 \displaystyle \int_{\partial \Omega}   \left(  (\C:\nabla \tilde {\mathbf u}).  {\mathbf n} \right).\tilde{\mathbf{ u}}  ~d{{\mathbf
x}}
=\displaystyle \int_{\partial \Omega} {\mathbf f}\Lambda_{\C}({\mathbf f}) d{\mathbf
x}
\end{eqnarray}

Now we introduce an orientation preserving, smooth and invertible  map  $F({\mathbf x})={\mathbf y}$ , on $\Omega,$ such that  $F({\mathbf x})={\mathbf x}$ for every ${\mathbf x} \in \partial\Omega.$ Then, we can rewrite $ \EuFrak{a_\C} ({\mathbf u},{\mathbf v})$ as

\begin{eqnarray}
 \EuFrak{a_\C} ({\mathbf u},{\mathbf v})=\displaystyle \int_{F(\Omega)} ( \C^F:\nabla_{\mathbf y}  {\mathbf u}):\nabla_{\mathbf y}{\mathbf v} ~d{\mathbf y},
\end{eqnarray}
where   $ \C^F$ is the  elasticity tensor of the transformed media $F(\Omega),$  obtained from $\C$ via $F,$ with components
\begin{eqnarray}
 \C^F_{ijkl} =\left( \det( \frac{d{\mathbf y}}{d{\mathbf x}})\right)^{-1} \displaystyle \sum_{p,q}\frac{\partial y_i}{\partial x_p}\frac{\partial y_k}{\partial x_q} \C_{pjql}
\end{eqnarray}

Hence for any ${\mathbf f}\in H^{\frac{1}{2}}(\partial \Omega),$ we chose a solution    $\tilde{\mathbf u}$ of  Eq. (\ref{eq:navier-general})  satisfying ${ \tilde{ \mathbf  u}}|_{\partial\Omega}={\mathbf f},$ so that  
\begin{eqnarray}
 \displaystyle \int_{\partial \Omega} {\mathbf f}\Lambda_{\C}({\mathbf f}) d{\mathbf
x}=  \EuFrak{a}_\C (\tilde{\mathbf u},\tilde{\mathbf u})= \EuFrak{a}_{\mathbb C^{F}} (\tilde{\mathbf u},\tilde{\mathbf u})= \displaystyle \int_{\partial \Omega} {\mathbf f}\Lambda_{\C^F}({\mathbf f}) d{\mathbf
x}.
\end{eqnarray}

Thus we have proved  the following property 
\begin{lemma} The linear operator $\Lambda$ is invariant under change of coordinates leaving $\partial\Omega$ pointwise invariant. That is, the quadratic forms $\Lambda_\C$  and $\Lambda_{\C^F}$  on the vector space $H^{\frac{1}{2}}(\partial \Omega),$ coincide.
\end{lemma}

\noindent
$\mathbf{Remark:}$ In a way similar to what happens in electric impedance tomography \cite{greenleaf,kohn2}, if $F( {\mathbf x})=  {\mathbf x}$ at   ${\partial \Omega},$ then the boundary measurements associated with $\C$ and $\C^F$ are identical (the change of variable does not affect the Dirichlet data.)

\subsection{Ordering elasticity tensors, Dirichlet data and  discussions }

Using formula (\ref{T-isotropic}), we can write in the isotropic homogeneous case, if ${\mathbf u}$ belongs to $\left[C^1(\Omega)\right]^n,$ 
\begin{eqnarray}\label{eq:div-strain}
\nabla. \sigma_1 = (\lambda+\mu)\nabla (\nabla .{\mathbf u}) +\mu\Delta {\mathbf u} =  (\lambda+\mu)\text{grad}(\text{div}~ {\mathbf u}) +\mu\Delta {\mathbf u}
\end{eqnarray}

Remark that for any two elasticity tensors, say $\C_1$ and $\C_2,$ that are not necessarily isotropic  nor homogeneous,  in $\Omega,$  if for every second order tensor $\xi$,  we have
\begin{eqnarray}\label{lem:first-inequality-tensors}
 (\C_1:\xi):\xi \leq (\C_2:\xi):\xi , \text{ ~then ~}  \Lambda_{\C_1}({\mathbf f}) \leq \Lambda_{\C_2}({\mathbf f}), 
\end{eqnarray}
 for all ${\mathbf f}$ in $\Hat  H^{\frac{1}{2}}(\partial\Omega).$

Due to the equality (in the case where $\C$ is isotropic homogeneous) 
\begin{eqnarray}(\C:\xi):{\xi'}&=&(\lambda+2\mu)\displaystyle\sum_i\xi_{ii}\xi_{ii}'+\lambda\displaystyle\sum_{i<j}\left(\xi_{ii}\xi_{jj}'+\xi_{jj}\xi_{ii}'\right) \nonumber\\
&+& \mu\displaystyle\sum_{i<j}\left[ \xi_{ij}(\xi_{ij}'+\xi_{ji}')+ \xi_{ji}(\xi_{ij}'+\xi_{ji}')\right],\end{eqnarray}
we have $(\C:\xi):{\xi'}=0,$ for all   tensors $\xi$  of order 2, whenever $\xi'$ is a skew-symmetric tensor of order 2.
Hence, we may only consider symmetric tensors of order 2. Moreover, if $\mu=0$, then as a quadratic form on the space of symmetric tensors of order 2, $\C$ becomes degenerate and has for kernel, the space of traceless symmetric tensors. More precisely, we have $(\C:\xi ):\xi =\lambda\left(\text{Trace}(\xi)\right)^2 . $  If $\mu>0,$ then $\C$ is non-degenerate (resp. definite positive) on the space of symmetric tensors of order 2, if and only if $n\lambda+2\mu$ is nonzero (resp. positive.)

\begin{lemma}\label{lem:second-inequality-tensors}
If $\C_1$ and $\C_2$ are two isotropic and homogeneous  elasticity tensors  in $\Omega\subset \R^n$ , with Lam\'e coefficients $\lambda_i$ and $\mu_i,$ $i=1,2,$ respectively. Then 
$(\C_1:\xi):\xi \leq (\C_2:\xi):\xi$  for every second order symmetric tensor $\xi$  in $\R^n,$ if and only if  $n\lambda_1+2\mu_1 \leq n\lambda_2+2\mu_2$ and  $\mu_1\leq\mu_2.$
\end{lemma}

\proof  For a symmetric tensor $\xi$ of order 2; we write $(\C_k:\xi ):\xi$ as
\begin{eqnarray}\label{eq:positifnesstensor}
(\C_k:\xi ):\xi =\frac{1}{n} (n\lambda_k+2\mu_k)\displaystyle\left(\sum_{i=1}^n\xi_{ii}\right)^2 + 4\mu_k\displaystyle\sum_{0\le i<j\le n}\xi_{ij}^2+ \frac{2\mu_k}{n}\displaystyle\sum_{0\le i<j\le n}\left(\xi_{ii}-\xi_{jj}\right)^2,\end{eqnarray}
 and hence
\begin{eqnarray}
(\C_2:\xi ):\xi -(\C_1:\xi ):\xi &=& \frac{1}{n} \left(n\lambda_2+2\mu_2-(n\lambda_1+2\mu_1)\right)\displaystyle\left(\sum_i\xi_{ii}\right)^2 \nonumber \\
&+& 4(\mu_2-\mu_1)\displaystyle\sum_{i<j}\xi_{ij}^2+ \frac{2}{n}(\mu_2-\mu_1)\displaystyle\sum_{i<j}\left(\xi_{ii}-\xi_{jj}\right)^2.\end{eqnarray}
\qed

Of course, the condition $\lambda_1\leq \lambda_2$ and $\mu_1\le\mu _2,$ is sufficient for $(\C_1:\xi):\xi \leq (\C_2:\xi):\xi$  to hold true for every second order symmetric tensor $\xi.$ But it is not necessary! For example, in dimension $n=2,$ we could choose $1.3=\lambda_1>\lambda_2=1.2$ and $1=\mu_1<\mu_2=1.5,$ but yet we have $4.6=2\lambda_1+2\mu_1<2\lambda_2+2\mu_2 = 5.4$ and hence, the above inequality is still true for all symmetric $\xi.$ 

\begin{lemma}\label{lem:third-inequality-tensors}
If $\C$ is the elasticity tensor of an isotropic homogeneous medium occupying the bounded domain $\Omega\subset\mathbb R^n,$ with Lam\'e Coefficients $\lambda$ and $\mu$, then, for every symmetric tensor $\xi$ of order 2, we have 
\begin{eqnarray} C_1(\lambda,\mu)||\xi ||^2\le (\C:\xi):\xi\le C_2(\lambda,\mu)||\xi ||^2, \end{eqnarray}
where the constants $C_1(\lambda,\mu)$ and $C_2(\lambda,\mu)$ are $C_1(\lambda,\mu):=\min\{n\lambda+2\mu,2\mu \}$ and  $C_2(\lambda,\mu)=\sup\{n\lambda+2\mu,2\mu \}.$ 
\end{lemma}
\proof
Let  $\xi$ be a symmetric tensor of order 2,  we can write $||\xi||^2$ as 
\begin{eqnarray} ||\xi||^2=\frac{1}{n}\displaystyle\left(\sum_i\xi_{ii}\right)^2 + 2\displaystyle\sum_{i<j}\xi_{ij}^2+ \frac{1}{n}\displaystyle\sum_{i<j}\left(\xi_{ii}-\xi_{jj}\right)^2,\end{eqnarray} 
 so using Eq. (\ref{eq:positifnesstensor}), we get $C_1(\lambda,\mu)||\xi ||^2\le (\C:\xi):\xi\le C_2(\lambda,\mu)||\xi ||^2.$

\qed

From now on, we will mostly be assuming that for the elasticity tensors $\C({\mathbf x}),$   there are two positive  constants $C_1$ and  $C_2$ such   $C_1||\xi ||^2\le (\C({\mathbf x}):\xi):\xi\le C_2||\xi ||^2$ for all symmetric tensor $\xi$  of order 2 on $\mathbb R^n$ . So the solution to Eq. (1), together with some boundary conditions, is well defined and unique.

Now we look for two isotropic homogeneous elasticity tensors with Lam\'e coefficients large enough (resp. small enough) in order to apply Inequality (\ref{lem:first-inequality-tensors}) and Lemma \ref{lem:second-inequality-tensors} to bound from above (resp. below) any given elasticity tensor that might be non-isotropic and non-homogeneous. Let us now consider the special case of isotropic homogeneous elasticity tensors with Lam\'e coefficients, say $\lambda'$ and $\mu',$ satisfying  $\lambda'=a\mu'\ge 0,$ where $a$ is a number. Without lost of generality, we take $a=2.$  We let $\C_{\lambda'}$ stand for the corresponding elasticity tensor,  so $\C_0$ will correspond to case $\lambda'=0.$  In this scheme, the Poisson coefficient $\nu:=\frac{\lambda}{2(\lambda+\mu)}$ will be equal to $1/3.$
 We also  have  
\begin{eqnarray}\label{eq:extremes}
(\C_{\lambda'}:\xi):\xi=\lambda'||\xi||^2+\lambda'\displaystyle\left(\sum_{i=1}^n\xi_{ii}\right)^2\ge \lambda '||\xi||^2.\end{eqnarray}
 So from Lemma \ref{lem:second-inequality-tensors}, for any isotropic homogeneous elasticity tensor $\C$ with Lam\'e coefficients $\lambda,$ $\mu,$ we will always have 
\begin{eqnarray} \label{eq:majoration}0= (\C_0:\xi):\xi \le (\C:\xi):\xi \le  (\C_{\lambda'}:\xi):\xi \end{eqnarray}
 as long as $\lambda'\ge \max\{n\lambda+2\mu,2\mu \}$ and  $ n\lambda+2\mu \ge 0.$
More generally, for any (not necessarily isotropic homogeneous) elasticity tensor $\tilde \C,$ such that $ (\tilde \C:\xi):\xi$ is  bounded, with nonnegative lower bound, for all  symmetric $\xi,$ we can choose some large enough $\lambda'$ so that inequality (\ref{eq:majoration}) still hold for $\tilde \C.$
 Inequality (\ref{lem:first-inequality-tensors}) then implies that 
\begin{eqnarray} \label{eq:majoration2}\Lambda_{0} -\Lambda_{\C}\le \Lambda_{\tilde \C} -\Lambda_{\C}\le  \Lambda_{\C_{\lambda'}} -\Lambda_{\C}\end{eqnarray}
 as long as $\lambda' $ is sufficiently large, where  $\Lambda_{0},\Lambda_{\C}, \Lambda_{\tilde \C},$ and $  \Lambda_{\C_{\lambda'}}$ are regarded as linear operators on $H^{\frac{1}{2}}(\partial\Omega).$

\subsection{Inclusions and cloaking}
In this section, we consider an isotropic homogeneous elastic medium, say, the ball $\Omega:=B({\mathbf x}_0,R_2)$  of radius $R_2$ centred at some point ${\mathbf x}_0,$ as above. For convenience,  we will take  ${\mathbf x}_0$  to be the origin of coordinates,  ${\mathbf x}_0={\mathbf 0}$  and $R_2=2.$ The elasticity tensor on $\Omega$ is $\C,$ with corresponding Lam\'e coefficients $\lambda$ and $\mu.$
For some  $\rho >0,$ we let ${B_\rho}=B({\mathbf 0},\rho)$ stand for the ball (disc for $n=2$) of radius $\rho$ centred at  ${\mathbf x_0}={\mathbf 0}$ . Let  $\C_\rho$  be some elasticity tensor in  ${B_\rho}=B({\mathbf 0},\rho).$ The tensor  $\C_\rho$ might be anything, e.g. fully anisotropic and non-homogeneous.
For a given tensor $\C,$ we call  $\Omega_{\C_\rho,\C}$ the ball  $\Omega:=B({\mathbf x}_0,R_2)$ filled with an elastic medium with elasticity tensor ${\mathbf A}_{\C_\rho,\C}$ defined by 
\begin{eqnarray}
{\mathbf A}_{\C_\rho,\C} = 
\left\{
\begin{array}{lr}
 \C  \text{ in } \Omega\setminus \overline{B_\rho} 
\\
  \C_{\rho}  \text{ in }  B_\rho
\end{array}
\right.
\end{eqnarray}
In all what follows, we assume that $  \C_{\rho}  $ is such that ${\mathbf A}_{\C_\rho,\C}$ is positif definite and bounded.

When $\C$ is isotropic homogeneous with Lam\'e coefficients  $\lambda,$ $\mu$ and $\C_\rho$ is of the type $\C_{\lambda'},$ as in (\ref{eq:extremes}), we simply write ${\mathbf A}(\lambda, \mu,\lambda') $  instead of  ${\mathbf A}_{\C_{\lambda'},\C},$ namely 
\begin{eqnarray}
{\mathbf A}(\lambda; \mu,\lambda') = 
\left\{
\begin{array}{lr}
 \C  \text{ in } \Omega\setminus \overline{B_\rho} 
\\
  \C_{\lambda'}  \text{ in }  B_\rho
\end{array}
\right.
\end{eqnarray}

For a given ${\mathbf  f}$ on $ H^{\frac{1}{2}}(\partial \Omega),$ we let  ${\mathbf u}_{{\C_\rho,\C}}$  stand for the solution to the boundary value problem  $
\nabla.\left( {\mathbf A}_{\C_\rho,\C}:{\nabla \mathbf u}_{{\C_\rho,\C}}\right)= {\mathbf 0}$ in $\Omega$  with  ${\mathbf u}_{{\C_\rho,\C}}= {\mathbf  f}$ on $\partial \Omega.$

From now on, we fix $\C$ and suppose it is isotropic homogeneous with Lam\'e coefficients $\lambda, \mu.$  

In practice, for the case $\C_\rho=\C_{\lambda'}$ and when $\lambda'=0,$
${\mathbf u}_{{\C_0,\C}}$ will then be the solution to 
  $ (\lambda+\mu)\text{grad(div}~{\mathbf u}_{{\C_0,\C}}) +\mu\Delta {\mathbf u}_{{\C_0,\C}}$ in $\Omega\setminus \overline{B_\rho}$ with $\left(\mathbb C:\nabla {\mathbf u}_{{\C_0,\C}}\right). {\mathbf n}=0$ on $\partial B_\rho$ and  ${\mathbf u}_{{\C_0,\C}}= {\mathbf  f}$ on $\partial \Omega.$

We define

\begin{eqnarray}\label{eq:inequaliy-Dirichlet-to-Neumann}
 \Lambda_{\C_\rho,\C} f:= \left(({\mathbf A}_{\C_\rho,\C}:\nabla{\mathbf u}_{\C_\rho,\C}).{\mathbf n}\right) |_{\partial \Omega}
\end{eqnarray}

Equation (\ref{eq:majoration2})  implies in particular that, for any bounded  $\C_\rho,$ we have the following
\begin{eqnarray} \label{eq:majoration3}\Lambda_{\C_0,\C} -\Lambda_{\C}\le \Lambda_{\C_\rho,\C} -\Lambda_{\C}\le \Lambda_{\C_{\lambda'},\C} -\Lambda_{\C}
\end{eqnarray}
 as long as $\lambda' $ is sufficiently large.

This means that we only need to control  $||\Lambda_{\C_0,\C} -\Lambda_{\C} ||$ and $||\Lambda_{\C_{\lambda'},\C} -\Lambda_{\C} ||.$
\medskip

\begin{figure}[h]\label{fig-dtn-asymptotic1}
\resizebox{160mm}{!}{\includegraphics{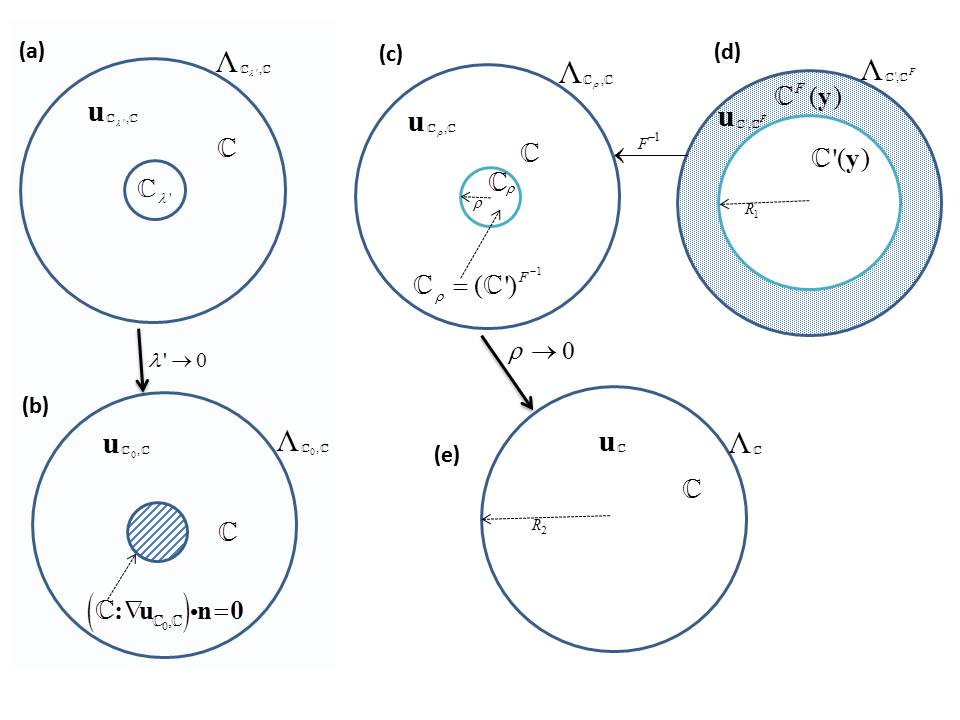}}
\caption{Schematic picture: (a) the domain $\Omega=B({\mathbf 0}, R_2),$ $R_2=2,$ with isotropic homogeneous elasticity tensor $\C$ in $\Omega\setminus B_\rho$ and $\C_{\lambda'}$ in $B_\rho.$  The solution to Equation (\ref{eq:navier-general}) is ${\mathbf u}_{\C_{\lambda' },\C}$  and the Dirichlet to Neumann map is denoted by $\Lambda_{\C_{\lambda'},\C}$. When $\lambda\to 0,$ we end up with (b) $\Omega$ with a spherical (circular, in 2D) hole of radius $\rho$ in its center and boundary conditions are
 $(\C:\nabla {\mathbf u}_{\C_0,\C}).{\mathbf n}=0,$ where $ {\mathbf u}_{\C_0,\C}$  is the solution to (\ref{eq:navier-general}). In (d) the domain $\Omega$ is made of an annulus $R_1 \leq ||{\mathbf y}||\leq R_2$  (circular ring, in 2D) in which the elasticity tensor $\C^F$ is obtained from the isotropic homogeneous $\C$ by the change of coordinates $F$; and the ball (resp. disc) $B(0,R_1)$  of radius $R_1=1,$ in which we put any tensor $\C'({\mathbf y})$.  In (e) the medium is isotropic homogeneous, with tensor $\C.$}
\label{figure}
\end{figure}

We have the following
\begin{proposition}  There exist positive constants $ C(\lambda,\mu)$   and $ C'(\lambda,\mu)$   such that
\label{andre}
\begin{eqnarray}
||\Lambda_{\C_0,\C} -\Lambda_{\C} ||\leq  C(\lambda,\mu) \rho^n \text{  and } ||\Lambda_{\C_{\lambda'},\C} -\Lambda_{\C}||\leq  C'(\lambda,\mu) \rho^n
\end{eqnarray}
when $\rho$ is sufficiently small
\end{proposition}
\proof
The proof is an adaptation of Proposition 1 in  p.12 of \cite{kohn2} to the elastostatic case. The complete derivation will be published elsewhere \cite{diatta-cooper-guenneau}. \qed

Consider an elastic medium occupying a ball  $\Omega=B_2:=B({\mathbf 0}, R_2)$ of radius $R_2$ about the origin, with  elastic tensor $\bar  A_{\C',C^F}$  of the form 
\begin{eqnarray}\label{eq:prooftomo1}
\bar  A_{\C',C^F}({\mathbf y}) =
\left\{
\begin{array}{lr}
\C'({\mathbf y})  ~~~ ~~~~~~~~~~~~~~~\text{ ~ if ~}~~~ ~~~{\mathbf y} \in B_1=B({\mathbf 0},R_1)
\\
\C_1^F({\mathbf y})  ,  ~~ ~~~~~~~~~~~~~~\text{ ~ if ~}~~~ ~~~   {\mathbf y} \in B_2\setminus  B_1.
\end{array}
\right.
\end{eqnarray}
The tensor $\C'({\mathbf y})$ is arbitrary (as above, we suppose it is such that $\bar A_{\C',C^F} ({\mathbf y})$ positive definite on symmetric tensors of order 2 and bounded) and it is the tensor in the region $B_1$ to be cloaked. We assume there are two positive constants $C_1$ and  $C_2$ such   $C_1||\xi ||^2\le (\bar  A_{\C',\C_1^F} :\xi):\xi\le C_2||\xi ||^2$ for all symmetric tensor $\xi$  of order 2 on $\mathbb R^n$ . (Hence the solution to Eq. (1), together with some boundary conditions, is well defined and unique.) The tensor  $\C_1$ is isotropic homogeneous with Lam\'e coefficients  $\lambda_1$  and $\mu_1.$ 

Below, we would like to show that the boundary measurements at $\partial \Omega$ are nearly the same as those in the situation where we have replaced $\C$ by another tensor which is everywhere equal to $\C_1$ inside the whole ball $\Omega,$ so it becomes an isotropic homogeneous medium. Let us call $\Lambda_1$ the Dirichlet-to-Neumann map corresponding to the latter situation, i.e.  the boundary measurements for  an isotropic homogeneous  elastic medium occupying  $\Omega,$ with constitutive  tensor everywhere equal to $\C_1$ inside the whole ball $\Omega.$

 As seen in Section \ref{chap:invariance-change-of-coordinates}, we have the equality 
  \begin{eqnarray}\label{eq:prooftomo1}
\Lambda_{\C',\C_1^{F}}  =\Lambda_{\C'^{F^{-1}},\C_1}. 
\end{eqnarray}
That is, the measurements at $\partial \Omega$ are the same, when $\Omega$ is occupied by two different media with tensors   $\bar  A_{\C',\C_1^F}$  and $A_{\C'^{F^{-1}},\C_1},$ respectively. Where $\Lambda_{\C'^{F^{-1}},\C_1} $ is the Dirichlet to Neumann map corresponding to the elasticity tensor
\begin{eqnarray}\label{eq:tomo_lambda}
A_{\C'^{F^{-1}},\C_1}
({\mathbf y}) =
\left\{
\begin{array}{lr}
{(\C')^{F^{-1}}}({\mathbf y})  ~~~ ~~~~~~~~~~~~~~~\text{ ~ if ~}~~~ ~~~{\mathbf y} \in B_\rho
\\
\C_1 ,  ~~ ~~~~~~~~~~~~~~\text{ ~ if ~}~~~ ~~~   {\mathbf y}  \in B_2\setminus B_\rho
\end{array}
\right.
\end{eqnarray}

Thus taking $\C_\rho=(\C')^{F^{-1}}$  and $\C=\C_1$ in  Inequality (\ref{eq:majoration3}),  we can apply it to $\Lambda_{\C_\rho,\C_1} -\Lambda_1$ followed by Proposition \ref{andre}, in order to control  by a factor of $\rho^n,$ the deviation $||\Lambda_{\C',\C_1^{F}} -\Lambda_1 ||  = ||\Lambda_{\C'^{F^{-1}},\C_1} -\Lambda_1 ||.$

\begin{definition}With the same notations as above, we say that the ball (disc, in 2D)  $B_1:=B({\mathbf 0},R_1)$ is nearly cloaked by $\C_1^F$ if  there exists a positive constant $C,$ such that for  any elasticity tensor $C'$ in $B_1,$  we have 
\label{almostcloak}
\begin{eqnarray}
||\Lambda_{\C'^{F^{-1}},\C_1}-\Lambda_{1} ||\leq C \rho^n,
\end{eqnarray}
for some positive constant $C.$
\end{definition}

Taking $F$ to be, for instance,
\begin{eqnarray}\label{eq:prooftomo1}
F({\mathbf x}) =
\left\{
\begin{array}{lr}
~~~~~~~~\frac{\mathbf  x}{\rho} ~~~ ~~~~~~~~~~~~~~~~~~\text{ ~ if ~}~~~ ~~~|| {\mathbf x} || \leq \rho
\\
\left( R_1+ \frac{R_2-R_1}{R_2-\rho}   ( ||  {\mathbf x} ||  - \rho) \right)\frac {  {\mathbf x} }{ || {\mathbf x} ||}, ~~~  \text{ ~ if ~}   \rho  \leq || {\mathbf x} || \leq R_2
\end{array}
\right.
\end{eqnarray}
where $R_2=2$ and $R_1=1,$ we have proved the following 
\begin{theorem} Consider an elastic medium occupying the shell $B_2\setminus B_1$ and having elasticity tensor $\C_1^F,$ obtained from $\C_1$ using the transformation  $F$ given in (\ref{eq:prooftomo1}).  If $\rho$ is small enough, then $B_1$ is nearly cloaked by $\C_1^F$ . 
\end{theorem}

\section{Numerical illustration with finite elements}
Let us show numerically that one can realize some cloaking as
described in proposition \ref{andre} with finite elements.
We setup the numerical model using the weak form of 
(\ref{eq:navier-general}) with nodal
elements. The forcing term ${\mathbf f}$
has a support on the boundary of a disc,
as shown in Figure \ref{figure}. The
effect of near  cloaking is obvious.

\begin{figure}[h!]\label{fig:figure-tomo-elasticity}
\resizebox{140mm}{!}{\includegraphics{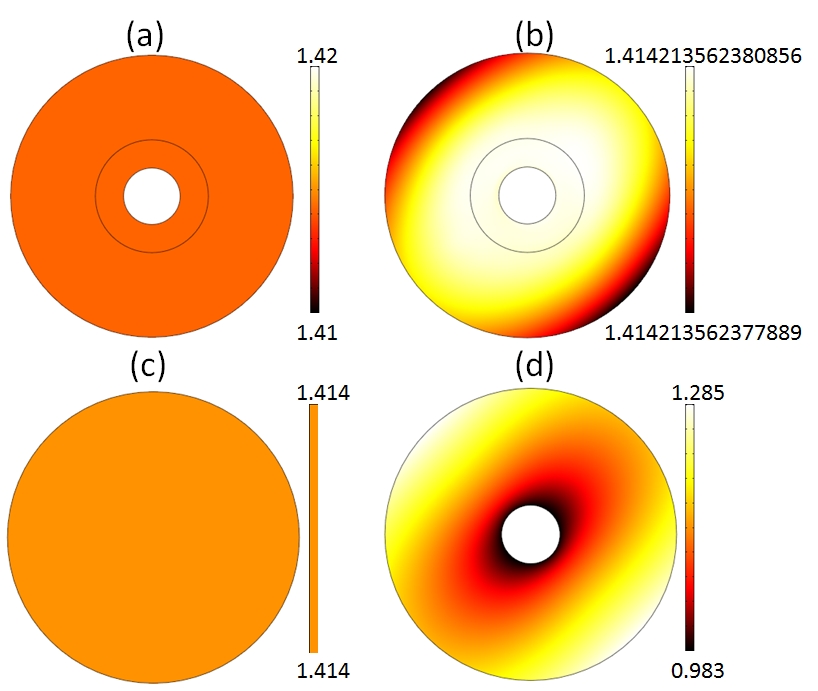}}
\caption{2D plots of magnitude of elastic field $\mid {\mathbf u}\mid=\sqrt{u_1^2+u_2^2}$
within a circular domain of radius $2$m with a normalized forced field ${\mathbf f}=(1,1)$ on its outer boundary:
(a) Domain with a stress-free circular hole (radius $0.2$m surrounded by a cloak (inner radius $0.2$m
and outer radius $0.4$m);
(b) Same as (a) with saturated color scale;
(c) Domain with no hole;
(d) Domain with a stress-free circular hole (radius $0.2$m). Note the much reduced range
for values in the color scales between (d) and (b). Plots in (a), (b) and (c) are nearly
identical, which is the effect of cloaking in elastic impedance tomography.}
\label{figure}
\end{figure}

\section{Conclusion}
In this article, we have introduced the concept of cloaking for elastic impedance tomography.
Some proposition has been derived which states that boundary measurements on a region
encompassing an elastic medium with and without a cloak nearly coincide. Some
numerical results are in agreement with this proposition.We hope this work will foster theoretical and numerical effort towards a better understanding  of zero and all frequencies elastic cloaking.

\section*{Acknowledgments}
The authors would like to thank Dr. S. Cooper for very helpful discussions and for validating the proof of Proposition 2.1.
The authors are thankful for European funding through ERC Starting Grant ANAMORPHISM.


\begin{thebibliography}{99}
\bibitem{willis}
G.W. Milton, M. Briane and J.R. Willis,
On cloaking for elasticity and physical equations with a transformation invariant form,
New J. Phys. {\bf 8}, 248 (2006).
\bibitem{norris}
A. Norris, Acoustic cloaking theory,
Proc. R. Soc. Lond. {\bf 464}, 2411 (2008).
\bibitem{brun}
M. Brun, S. Guenneau and A.B. Movchan,
Achieving control of in-plane elastic waves,
Appl. Phys. Lett. {\bf 94},
061903 (2009).
\bibitem{pendry}
Pendry, J.B., Schurig, D. and Smith, D.R. 2006
Controlling Electromagnetic Fields, Science {\bf 312}, 1780
\bibitem{leonhardt}
Leonhardt, U. 2006
Optical Conformal Mapping,
Science {\bf 312}, 1777
\bibitem{kohn1}
R.V. Kohn, and M. Vogelius 1984, Determining conductivity by boundary measurements, Comm.
Pure and Appl. Math. {\bf 37}, 289-298
\bibitem{lee}
J. Lee, and G. Uhlmann 1989, Determining anisotropic homogeneous real-analytic conductivities by boundary
measurements, Comm. Pure and Appl. Math. {\bf 42}, 1097-1112
\bibitem{greenleaf}
Greenleaf, A., Lassas, M. and Uhlmann, G. 2003
On nonuniqueness for Calderon's inverse problem,
 Math. Res. Lett. {\bf 10}, 685-693
\bibitem{kohn2}
R.V. Kohn, H. Shen, M.S. Vogelius, and M.I. Weinstein 2008
Cloaking via change of variables in electric impedance tomography,
Inverse Problems {\bf 24}, 015016
\bibitem{cummer}
S.A. Cummer and D. Schurig,
One path to acoustic cloaking,
New J. Phys. {\bf 9}, 45 (2007).
\bibitem{chan}
H. Chen and C.T. Chan,
Acoustic cloaking in three dimensions using acoustic metamaterials,
Appl. Phys. Lett. {\bf 91}, 183518 (2007).
\bibitem{fink}
M. Fink and C. Dorme 1997,
Aberration correction in ultrasonic medical imaging with time-reversal techniques,
International Journal of Imaging Systems and Technology,
Special Issue: Acoustical Tomography {\bf 8}(1),
110-125.
\bibitem{diatta-guenneau} A. Diatta and S. Guenneau, 
Non-singular cloaks allow mimesis, 
Journal of Optics {\bf 13} (2011), no.2, 024012-024022. 

\bibitem{diatta-cooper-guenneau} A. Diatta, S. Cooper and S. Guenneau, Work in preparation.
\end{thebibliography}
\end{document}